\documentclass[11pt]{article}

\pdfoutput=1

\usepackage[makeroom]{cancel}
\usepackage{color}
\usepackage{graphicx}
\usepackage{amsmath}
\usepackage{amssymb}
\usepackage{xspace}
\usepackage[small]{subfigure}
\usepackage[numbers,compress]{natbib}
\usepackage[hyperfootnotes=false]{hyperref}
\usepackage{tcolorbox}

\newlength{\fighskip} \fighskip=2pt
\newlength{\figvskip} \figvskip=3pt

\newcommand*{\figbox}[2]{{
  \def\figscale{#1}
  \def\arraystretch{0.8}
  \arraycolsep=0pt
  \begin{array}{c}
    \vbox{\vskip\figscale\figvskip
      \hbox{\hskip\figscale\fighskip
        \includegraphics[scale=\figscale]{#2}}}
  \end{array}}}

\usepackage{mciteplus} 
\usepackage{dcolumn}
\usepackage{bm}
\usepackage{verbatim}
\usepackage{amscd}
\usepackage{amsfonts}
\usepackage{setspace}
\usepackage{amsthm}
\usepackage{enumerate}
\usepackage{mathtools}

\begin{document}

\title{\bf 
Remarks on Black Hole Complexity Puzzle
}
\author{
Beni Yoshida\\ 
{\em \small Perimeter Institute for Theoretical Physics, Waterloo, Ontario N2L 2Y5, Canada} }
\date{}

\maketitle

\begin{abstract}
Recently a certain conceptual puzzle in the AdS/CFT correspondence, concerning the growth of quantum circuit complexity and the wormhole volume, has been identified by Bouland-Fefferman-Vazirani and Susskind. 
In this note, we propose a resolution of the puzzle and save the quantum Extended Church-Turing thesis by arguing that there is no computational shortcut in measuring the volume due to gravitational backreaction from bulk observers. A certain strengthening of the firewall puzzle from the computational complexity perspective, as well as its potential resolution, is also presented. 
\end{abstract}

\vspace{-0.7\baselineskip}
\section{Introduction}

Recently a certain conceptual puzzle in the AdS/CFT correspondence has been identified by Bouland-Fefferman-Vazirani (BFV)~\cite{Bouland19} and Susskind~\cite{Susskind20}. The puzzle is concerned with the growth of quantum circuit complexity in boundary CFTs and its dual on the bulk; the growth of the wormhole volume. It is widely believed that finding the quantum circuit complexity of a quantum state is generically a difficult computational problem. Thus, there should be no efficient way of computing the complexity of the boundary CFT wavefunction. But the wormhole volume is a macroscopic quantity which appears to be easily measurable by bulk observers. This would suggest that a certain physical phenomenon in quantum gravity may not be efficiently simulated on a quantum computer, violating the quantum Extended Church-Turing (qECT) thesis~\cite{Deutsch:1985aa}. 

In this note, we propose a resolution of this puzzle. Any attempt to see the black hole interior by an observer always introduces significant gravitational backreaction to the underlying geometry, and naive predictions on the black hole interior based on effective bulk descriptions break down. 
This invalidates protocols that would appear to measure the volume/complexity efficiently on the bulk, and thus the qECT thesis remains valid. 
To demonstrate this point explicitly, we discuss how to construct interior entangled partner operators with perturbations by a bulk observer taken into account. 
Namely, we claim that the construction of interior operators changes dynamically due to backreaction from an infalling observer.
This conclusion follows from a certain quantum information theoretic theorem which relates quantum information scrambling, as quantified by OTOCs, to the decoupling phenomenon. 
The theorem suggests that the infalling observer's backreaction disentangles the outgoing Hawking mode from the other side of the black hole. 
The bulk interpretation of the decoupling phenomena is deduced by considering the gravitational backreaction from the infalling observer herself. 
We also critically comment on a proposal by Susskind for resolving the complexity puzzle via the black hole complementarity~\cite{Susskind20}. 

Susskind derived his version of the complexity puzzle by relying on a certain hypothesis that underpins the $\text{ER}=\text{EPR}$ proposal for the firewall puzzle~\cite{Maldacena13}; two observers from the opposite sides of a two-sided AdS black hole will see each other inside the black hole even though two boundaries are not coupled. Our main result directly negates this hypothesis and the $\text{ER}=\text{EPR}$ proposal, suggesting that two observers will \emph{not} be able to see each other. While this conclusion is naturally expected from the locality of the boundary quantum mechanics, the important challenge is to understand how the naive bulk prediction, which led to a misunderstanding in previous works, breaks down due to backreaction from an infalling observer. Detailed discussions and concrete counterarguments against the $\text{ER}=\text{EPR}$ proposal have been presented in~\cite{Beni19, Beni19b}, which will be reiterated in this note. 

One crucial element in our counterargument against the $\text{ER}=\text{EPR}$ proposal is the surprising fact that one can save an infalling observer, who jumped into a black hole, via a quantum operation which is strictly localized on one side of the black hole~\cite{Beni19, Beni19b}. 
Indeed, this is the physical interpretation of the reconstruction protocol for the interior partner operator where the infalling observer returns to the exterior together with the partner mode. 
The recovery procedure extracts the infalling observer to the exterior even if a perturbation (be it unitary or non-unitary) is added on the other side of the black hole. 
This observation provides a strong evidence that the experience of the infalling observer was not affected by a perturbation on the other side. 

This note is organized as follows. In Section~\ref{Sec:Complexity} and~\ref{Sec:Puzzle}, we provide a brief review of the black hole complexity puzzle. 
In Section~\ref{Sec:ER=EPR}, we revisit previous proposals, such as the $\text{ER}=\text{EPR}$ proposal, critically.
In Section~\ref{Sec:Backreaction} and~\ref{Sec:bulk}, we discuss the backreaction from the infalling observer from the boundary and bulk perspectives respectively. 
In Section~\ref{Sec:Universality}, we argue that the disentangling phenomena is universal and occurs no matter how the infalling observer jumps into a black hole.  
In Section~\ref{Sec:Measurement}, we discuss whether performing a measurement on the other side of a black hole affects the infalling observer or not. 
In Section~\ref{Sec:Resolution}, we present the resolution of the black hole complexity puzzle. We also present a refined interpretation of a certain thought experiment due to Hayden and Preskill, concerning a possible encounter inside a one-sided black hole. 
In Section~\ref{Sec:extend}, we present a certain strengthening of the firewall puzzle from the perspective of the qECT thesis.
In Section~\ref{Sec:Discussion}, we conclude with brief discussions.
In Appendix~\ref{Sec:misc}, we present answers to some of the frequently asked questions.

\section{Complexity}\label{Sec:Complexity}

The quantum Extended Church-Turing (qECT) thesis is the following belief/physical principle~\cite{Deutsch:1985aa}: \vspace{0.2\baselineskip}

\emph{
-- All the physical processes that obey fundamental laws of physics, including quantum gravity, are efficiently simulable on a quantum computer. 
}
\vspace{0.2\baselineskip}

A thesis is a proposal for the sake of argument. It is not formulated as a precise mathematical conjecture and its correctness remains to be verified. If the qECT thesis is wrong, one could potentially speed up quantum computation by using quantum gravity effects~\footnote{It is implicitly assumed that a physical system can be described in a finite-dimensional Hilbert space. }. The AdS/CFT correspondence is a prime setup to test the qECT thesis as it posits that all the physical phenomena in the bulk quantum gravity are encoded in boundary quantum mechanical systems and can be efficiently simulated on a quantum computer.

To derive the puzzle, we will need the contraposition of the qECT thesis within the AdS/CFT correspondence.
\vspace{0.2\baselineskip}

\emph{
-- Any physical processes that cannot be efficiently simulated in boundary CFTs cannot be efficiently simulated in bulk AdS quantum gravity. 
}
\vspace{0.2\baselineskip}

To make the logic of this note clear, it is worth representing the statement of the qECT thesis schematically:
\begin{center}
\begin{tabular}{ c c c }
 bulk &      & boundary \\  \hline
 easy & $\rightarrow$ & easy \\  
 difficult & $\leftarrow$ & difficult   
\end{tabular}
\end{center}
Since the bulk consists both of quantum mechanics and gravity, it is reasonable to think that the bulk is stronger than (or equal to) the boundary in terms of the computational power~\footnote{If some problem is easy on the boundary, one can simply bring a quantum computer to the bulk and solve it without using gravity. Hence, ``easy $\leftrightarrow$ easy'' and ``difficult $\leftrightarrow$ difficult''.}. The qECT thesis says that the AdS gravity does not provide any quantum computational speedup. 

\vspace{0.5\baselineskip}

The central object of interest is the complexity of boundary wavefunctions in the AdS/CFT correspondence. The quantum circuit complexity $\mathcal{C}$ of a quantum state $|\psi\rangle$ is the minimal number of few-body quantum gates that are required to create $|\psi\rangle$ from some simple reference state such as $|0\rangle^{\otimes n}$. While the precise definition of $\mathcal{C}$ depends on choices of elementary gate sets and other details, these subtleties will not be essential in demonstrating and resolving the complexity puzzle.

Recall that the two-sided eternal AdS black hole is conjectured to be dual to the thermofield double (TFD) state $|\text{TFD} \rangle \propto \sum_{i}e^{-\beta E_i}|\psi_i\rangle_{L} \otimes |\psi_i\rangle_{R}$. Here we will focus on black hole geometries arising from time-evolutions of $|\text{TFD}\rangle$ under weak perturbations that can be treated as gravitational shockwaves. We will assume that the thermodynamic properties of the black hole, such as the temperature, do not change by such perturbations. The following conjecture was proposed~\cite{Stanford:2014aa}:
\vspace{0.2\baselineskip}

\emph{-- The quantum circuit complexity $\mathcal{C}$ of boundary CFT wavefunction is dual to the wormhole volume $\mathcal{V}$ in the maximal volume slice.}

\vspace{0.2\baselineskip}
In other words, $\mathcal{C} \approx \mathcal{V}$ in some appropriate normalization.

\vspace{0.5\baselineskip}

\begin{figure}
\centering
(a)\includegraphics[width=0.28\textwidth]{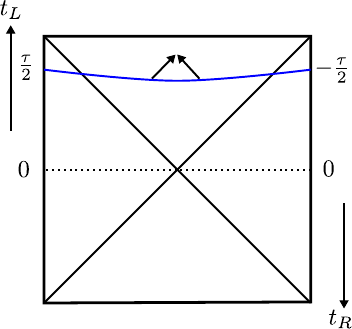} \
(b)\includegraphics[width=0.28\textwidth]{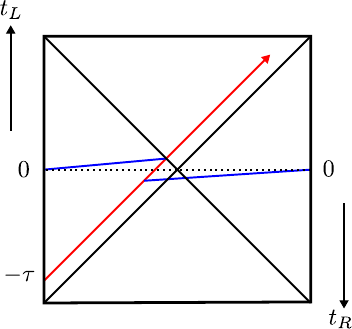} \
(c)\ \includegraphics[width=0.32\textwidth]{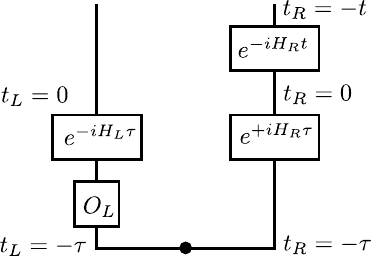}
\caption{(a) A maximal volume slice. A pair of bulk observers in the interior are shown. (b) A slice with a shock from the past $t_{L}=-\tau$. (c) A circuit representation of a shockwave geometry. The horizontal line represents the TFD state.
}
\label{fig-complexity-set}
\end{figure}

The $\mathcal{C} \approx \mathcal{V}$ conjecture has passed several non-trivial tests in a qualitative sense. To gain some insight, consider a time-evolved state $|\text{TFD}(t) \rangle \equiv ( e^{-iHt}\otimes I) |\text{TFD} \rangle$. It is naturally expected that the complexity of $|\text{TFD}(t) \rangle$ linearly increases in $t$ when $H$ is chaotic. While this statement is unproven, there is a large body of supporting evidence from various perspectives, such as quantum complexity theory~\cite{Aaronson16} and pseudorandomness~\cite{Roberts:2017aa, Brandao19}. On the bulk, consider the time-evolved TFD state $|\text{TFD}(t)\rangle = |\text{TFD}(\frac{t}{2}, - \frac{t}{2})\rangle$ and its dual wormhole in the maximal volume slice. Here we used the fact that $|\text{TFD}(t) \rangle = |\text{TFD}(t_{L},t_{R}) \rangle  \equiv (e^{-iHt_{L}}\otimes e^{+iHt_{R}})|\text{TFD} \rangle$, where $t=t_{L}-t_{R}$. The volume $\mathcal{V}$ indeed grows roughly linearly in time $t$ for an exponentially long time, as in Fig.~\ref{fig-complexity-set}(a). Hence, after some proper normalization, we expect to have $\mathcal{C}\approx \mathcal{V}$. Another justification of $\mathcal{C} \approx \mathcal{V}$ is obtained by invoking the tensor network representation of the spacetime~\cite{Pastawski15b}.
\vspace{0.5\baselineskip}

In demonstrating the black hole complexity puzzle, we will be interested in the computational difficulty of determining $\mathcal{C}$ from $|\psi\rangle$, not $\mathcal{C}$ itself. Suppose that some unknown wavefunction $|\psi\rangle$ is given. Without prior knowledge on how $|\psi\rangle$ was prepared, determining its quantum circuit complexity $\mathcal{C}$ is a difficult computational problem. In the most generic setting of the problem, we could at best try to apply quantum gates to see if $|\psi\rangle$ returns to some simple state. Since there are $\exp\big(O(\mathcal{C})\big)$ nearly-orthogonal states with complexity $\mathcal{C}$, the complexity of estimating $\mathcal{C}$ is expected to be $\sim\exp(\mathcal{C})$. 
\vspace{0.2\baselineskip}

\emph{
-- The quantum computational complexity of estimating $\mathcal{C}$ is $\sim \exp(\mathcal{C})$ in general, and thus is a difficult computational problem.
}
\vspace{0.2\baselineskip}

While unproven, there are quantum computational complexity theoretic arguments supporting this conjecture, see~\cite{Bouland19} for instance. 

\vspace{0.5\baselineskip}

Let us pause for a technical comment. If one knows the Hamiltonian $H$ a priori, it is not difficult to estimate the complexity of $|\text{TFD}(t)\rangle$ as one can undo the time-evolution and see if the state returns to $|\text{TFD}(0)\rangle$. To circumvent this counterargument, BFV considered an ensemble of wavefunctions generated by insertions of multiple randomly-chosen shockwaves $\{O_{1},\ldots,O_{m}\}$: 
\begin{align}
(I \otimes e^{-iH_{R}t_{m} } O_{m} e^{-iH_{R}t_{m-1} } \cdots  e^{-iH_{R}t_{1} }O_{1} )|\text{TFD}\rangle.
\end{align}
They provided an argument suggesting that it is indeed difficult to estimate $\mathcal{C}$ even if one knows $H$ since it is difficult to distinguish the above ensemble from some random ensemble of much higher complexity. Here it is important to assume that the choice of $\{O_{1},\ldots,O_{m}\}$ remains unknown to the boundary observer. Unless otherwise stated, we shall focus on $|\text{TFD}(t)\rangle$ without insertion of multiple shockwaves for simplicity of discussion. 
\vspace{0.5\baselineskip}

Combining the above conjecture with another one $\mathcal{C}\approx \mathcal{V}$, we arrive at the following conjecture:
\vspace{0.2\baselineskip}

\emph{
-- Estimating $\mathcal{V}$ from generic boundary CFT wavefunctions is a difficult computational problem.
}
\vspace{0.5\baselineskip}

Below is a side comment. There is a certain heuristic relation between two-point correlation functions and the complexity of estimating $\mathcal{C}$. Two-point correlation functions between two sides of a black hole decay exponentially as the separation increases. This suggests that measurement of two-point functions by bulk observers, which would go as $\approx e^{- \gamma \mathcal{C}}$, can give an estimate of the wormhole length where $\gamma$ is some positive constant. When the complexity $\mathcal{C}$ is high, the measurement outcome is too weak, and one would need $\approx e^{+ \gamma \mathcal{C}}$ samples to make a reliable measurement of two-point functions. This is consistent with the fact that the complexity of estimating $\mathcal{C}$ is conjectured to be $\approx e^{\gamma' \mathcal{C}}$ for some constant $\gamma'$. Turing the argument around, the exponential growth of the complexity of estimating $\mathcal{C}$ would suggest the exponential (not polynomial) decay of two-point functions with $\gamma \gtrapprox \gamma'$. Interestingly, bulk field theories can put some restrictions on the complexity of estimating $\mathcal{C}$~\footnote{It is also interesting to note that two-point correlation functions at late times measure spectral form factors~\cite{Cotler:2017aa}. These contain linearly growing contribution, much like the wormhole volume~\cite{Cotler:2017ab}. }. If the bulk theory is somehow gapless with polynomially decaying correlation functions, the complexity of estimating $\mathcal{C}$ would be at most $\text{poly}(\mathcal{C})$.

\section{Puzzle}\label{Sec:Puzzle}

In the light of the qECT thesis, estimating the wormhole volume $\mathcal{V}$ should also be computationally difficult since estimating $\mathcal{C}$ is difficult. This is, however, strange. The wormhole volume is roughly proportional to the length of the wormhole which is a simple macroscopic quantity that appears to be easily computed or measured. This would suggest that estimating $\mathcal{V}$ may actually be a computationally tractable problem for bulk observers, violating the qECT thesis. 

BFV sharpened this observation by proposing a certain protocol to measure the wormhole volume by using multiple observers who live on multiple copies of the system. Imagine that one somehow populates the black hole interior with many observers along the wormhole, and see if pairs of observers can send signals to one another before reaching the singularity (Fig.~\ref{fig-complexity-set}(a)). The number of successful signal transmissions will be a coarse estimate of the wormhole volume. 

One potential problem in the BFV protocol is the use of multiple observers who are causally disconnected. Another issue is that to place observers in the interior, one would need to know the shape of the wormhole a priori, which implicitly assumes prior knowledge of $\mathcal{V}$. At the same time, however, the description of the protocol in~\cite{Bouland19} is rather brief, and we are unsure if we have addressed its full intent~\footnote{Perhaps a simpler protocol would be to prepare a pair of observers from two sides, without telling them what $H$ is, and let them meet inside the black hole and compare their watches. Unfortunately, this protocol does not work as we will discuss later. }.

\vspace{0.5\baselineskip}

Susskind presented another version of the puzzle which is free from these problems. Suppose that some boundary wavefunction $|\psi(t)\rangle$ undergoes time-evolution. Since it is difficult to estimate the complexity of $|\psi(t)\rangle$, it should be also difficult to judge if $\mathcal{C}(t)$ is increasing or decreasing at a given time. Combining this observation with the $\mathcal{C}\approx \mathcal{V}$ conjecture, the following conjecture is obtained.
\vspace{0.2\baselineskip}

\emph{
-- Judging if $\mathcal{V}(t)$ is increasing or decreasing from boundary CFT is computationally difficult.
}
\vspace{0.2\baselineskip}

To demonstrate the puzzle based on this observation, let us look at two classes of boundary wavefunctions. First, consider the time-evolution of the TFD state $|\Psi(t)\rangle \equiv |\text{TFD}(0,-t)\rangle$ at $t_{L}=0$ and $t_{R}=-t$. As we have seen, $\mathcal{C}(t)$ of $|\Psi(t)\rangle$ increases since $\mathcal{V}(t)$ increases in time (Fig.~\ref{fig-complexity-set}(a)). 

Next, consider another wavefunction at $t_{L}=0$ and $t_{R}=-t$ where, at some negative time $t_{L}=-\tau$, a small perturbation is applied (Fig.~\ref{fig-complexity-set}(b)):
\begin{align}
|\Phi(t) \rangle \equiv (e^{-iH\tau} \otimes I_{R}) (O_{L}\otimes I_{R})|\text{TFD}(-\tau,-t)\rangle.
\end{align}
It has been pointed out that $\mathcal{C}(t)$ of $|\Phi(t) \rangle $ actually decreases for $\tau \gtrapprox t$. On the bulk, this can be verified by directly computing the volume $\mathcal{V}$, see~\cite{Stanford:2014aa} for details. On the boundary, this decrease can be easily seen by drawing a quantum circuit diagram as in Fig.~\ref{fig-complexity-set}(c). Here we have prepared the initial state $|\Phi(0)\rangle$ by inserting a perturbation on the thermofield double state $|\text{TFD}(0)\rangle = |\text{TFD}(-\tau,-\tau)\rangle$ at $t_{L}=t_{R}=-\tau$, and time-evolving both sides to $t_{L}=t_{R}=0$. We see that time-evolution of $e^{-i H_{R}t}|\Phi(0)\rangle$ on the right-hand side cancels the $e^{i H_{R}\tau}$, and hence the complexity $\mathcal{C}(t)$ actually decreases until $t = \tau$. 

The takeaway from the above analysis is that complexity of wavefunctions can decrease or increase depending on whether a perturbation is applied on the left side or not. This is, however, strange from the bulk observer's perspective. An observer simply needs to jump into the black hole and see if she will be hit by a gravitational shockwave which lets her allow to judge the sign of $\frac{d\mathcal{C}}{dt}$. Once again, the qECT thesis seems to be violated. 

\vspace{0.5\baselineskip}

Several flaws in the above arguments can be immediately identified. Introducing a bulk observer behind the horizon corresponds to the high complexity quantum operation on the boundary. Namely, when approaching near the horizon, an infalling observer herself induces significant gravitational backreaction to the underlying geometry. These effects often invalidate bulk effective descriptions, and hence should be examined carefully, which is exactly what we will do in the next section. Also, it is unclear why an infalling observer from the right-hand side can see a shockwave from the left-hand side while two sides are not coupled. Later we shall argue that an observer will not be able to see the shockwave due to the backreaction caused by the observer herself. 

In the remainder of the note, we argue that measuring $\mathcal{V}$ or $\frac{d\mathcal{V}}{dt}$ on the bulk is not necessarily a computationally easy task by examining the effect of backreaction by infalling observers.

\vspace{0.5\baselineskip}

For convenience of readers, we summarize the argument of the complexity puzzle~\footnote{BFV's argument is a bit more involved and considers the complexity of the holographic dictionary. We will comment on it in Section~\ref{Sec:Discussion}.}. 
\vspace{-0.2\baselineskip}
\begin{enumerate}[\ \ 1).]
\item On the boundary, quantum circuit complexity $\mathcal{C}$ (or $\frac{d\mathcal{C}}{dt}$) is believed to be not efficiently computable. 
\item On the bulk, the wormhole volume $\mathcal{V}$ (or $\frac{d\mathcal{V}}{dt}$) seems to be efficiently measurable by infalling observers. 
\item The wormhole volume $\mathcal{V}$ and the quantum circuit complexity $\mathcal{C}$ are roughly proportional to each other. 
\item The qECT says that, as $\mathcal{C}$ (or $\frac{d\mathcal{C}}{dt}$) is not efficiently computable, $\mathcal{V}$ (or $\frac{d\mathcal{V}}{dt}$) should not be efficiently measurable either.
\end{enumerate}
\vspace{-0.2\baselineskip}
We will argue that 2) is wrong. Here 4) is the statement of the qECT thesis while 1) and 3) are widely-accepted conjectures.

\section{Interlude}\label{Sec:ER=EPR}

The underlying problem in Susskind's (and implicitly BFV's) argument is the following hypothesis.

\vspace{0.2\baselineskip}
\emph{-- In a two-sided black hole, an infalling observer from one side can see a signal from the other side in the black hole interior even when two sides of the black hole are not coupled (?)}
\vspace{0.2\baselineskip}

In this section, we revisit this hypothesis critically~\footnote{This puzzle was identified in~\cite{Marolf:2012aa} which asked why two observers from the two sides appear to be able to meet inside the black hole. Our resolution of this rendezvous puzzle, however, differs from the scenario proposed in~\cite{Marolf:2012aa}.}. 
\vspace{0.5\baselineskip}

\begin{figure}
\centering
\includegraphics[width=0.30\textwidth]{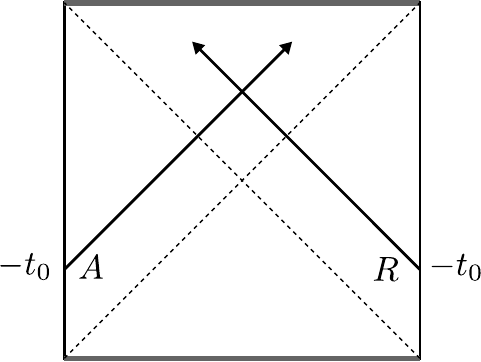}
\caption{A two-sided AdS black hole. The causal structure appears to suggest that two signals $A,R$ will encounter each other inside the black hole. This, however, leads to non-locality on the boundary.
}
\label{fig-meeting}
\end{figure}

From the bulk viewpoint, this hypothesis might appear reasonable. Indeed, trajectories of the infalling observer $A$ from one side and a perturbation $R$ from the other side appear to intersect with each other inside the black hole (Fig~\ref{fig-meeting}). Many of the previous proposals for resolutions of the firewall puzzle, including the $\text{ER}=\text{EPR}$ proposal, the quantum circuit complexity proposal~\cite{Harlow:2013aa} and the state-dependence proposal~\cite{Papadodimas:2013aa}, explicitly or implicitly assume this hypothesis. From the boundary viewpoint, however, this is rather strange. Suppose that $A$ and $R$ from opposite sides will encounter each other inside the black hole. Furthermore, suppose that $R$ can somehow influence $A$'s infalling experience as advocated in the $\text{ER}=\text{EPR}$ proposal. This would suggest that degrees of freedom on opposite sides of the boundary somehow interact with each other even though two sides are not coupled. Hence, the above hypothesis leads to an apparent non-locality on the boundary. 

This non-locality problem is well known. There are two logically reasonable approaches to resolve this problem. The first approach is to trust the bulk prediction: 

\vspace{0.2\baselineskip}
\emph{-- An infalling observer will see a signal from the other side. Hence, the locality in the boundary quantum mechanics is violated. }
\vspace{0.2\baselineskip}

And the second approach is to trust the boundary quantum mechanics: 

\vspace{0.2\baselineskip}
\emph{-- An infalling observer will not see a signal from the other side. Hence, the causal structure, implied by the Penrose diagram, becomes invalid.}
\vspace{0.2\baselineskip}

Note that both approaches deviate from standard arguments and thus require justifications. In the past, the first approach has been favoured in the community without much scrutiny, perhaps due to the celebrity effect. In this note, we negate the first approach and advocate for the second approach. In the remainder of this section, we present a certain counterargument against the first approach. 
\vspace{0.5\baselineskip}

The first approach requires one to reconcile an apparent non-locality resulting from the encounter with the locality of the boundary quantum mechanics. It has been previously argued that an apparent non-locality may not be detectable from the boundary observer. While there are a variety of such arguments, the following line of reasoning is commonly used. To verify the non-locality from the boundary quantum mechanical viewpoint, the boundary observer may hope to confirm that two observers have actually met inside the black hole. It appears, however, difficult to verify such an encounter that would happen behind the horizon. Hence, an apparent non-locality may not lead to any observable inconsistency in the boundary quantum mechanics. 

The crucial issue in this reasoning is whether the boundary observer can verify the encounter or not. If so, whether the verification can be performed efficiently is also an important question. Recent developments concerning the Hayden-Preskill thought experiment and traversable wormholes have shed new light on the above line of reasoning~\cite{Yoshida:2017aa, Gao:2017aa}. One possible way to verify the encounter is to extract infalling observers from the black hole interior and ask if they have actually met or not. One can extract infalling observers by implementing the Hayden-Preskill recovery protocol or the traversable wormhole protocol when the system is prepared in a thermofield double state. While these protocols require significant fine-tuning, quantum circuit complexity of their implementation is relatively low and hence is physically implementable in principle.

Thus, a naturally arising question is whether these recovery procedures lead to any detectable non-locality in the boundary quantum mechanics.
Recall that these recovery protocols require one to non-locally couple two sides of the black hole. 
Then, since an apparent non-locality is verified by non-locally coupling two sides, these recovery procedures do not lead to any violation of the locality from the perspective of the boundary observer~\footnote{We are unsure who originally proposes this perspective while we noticed that Maldacena advocates for it in his talk available at http://pirsa.org/19040126/.}. Hence, the first approach does not run into trouble if the following statement is true.

\vspace{0.2\baselineskip}
\emph{-- 
Whether two observers from opposite sides have actually encountered each other inside the black hole cannot be verified from the outside unless two sides are coupled (?)
}
\vspace{0.2\baselineskip}

\vspace{0.5\baselineskip}

The above line of reasoning for the first approach is unfortunately incorrect. It turns out that it is possible to extract an infalling observer from the black hole interior via a certain procedure which is strictly localized on one side of the black hole~\cite{Beni19b}. 
This procedure is closely related to how interior partner operators are constructed in the presence of the infalling observer and will be briefly reviewed in the next section. The upshot is that after performing this recovery procedure, one can bring the infalling observer $A$ back to the exterior and also distills a qubit that is entangled with the outgoing Hawking mode. This enables one to obtain the expression of the interior entangled partner mode explicitly. 

Hence, one can extract an infalling observer $A$ from the black hole interior by using the one-sided recovery protocol. So let us ask her if she has seen a signal $R$ from the other side of the black hole. While deriving $A$'s experience from a direct bulk calculation is beyond the scope of our discussion, it is possible to deduce that she never encountered $R$. This one-sided recovery procedure is strictly localized on $A$'s side of the black hole and succeeds even if perturbations are added on the other side. This suggests that $A$'s infalling experience will not be influenced by $R$ at all, favouring the no encounter scenario. It is worth emphasizing that the one-sided recovery works regardless of how the black hole is initially entangled with the early radiation and works for a one-sided black hole too. From these observations, we deduce that the infalling observer $A$ will never see a signal $R$ from the other side as long as two sides are not coupled.

\vspace{0.5\baselineskip}

We have presented an argument that negates the first approach. In the next two sections, we will present an argument supporting the second approach by considering the effect of backreaction from the infalling observer and finding interior partner operators. Results in the preceding sections will also provide additional counterarguments against the $\text{ER}=\text{EPR}$ proposal for the firewall puzzle.

\section{Backreaction}\label{Sec:Backreaction}

Let us return to the central issue of this note. The main goal of this section and the next is to discuss the backreaction from an infalling observer on the black hole interior geometries. Our strategy is to find interior partner modes that are entangled with outgoing modes in boundary CFTs. By studying how entanglement structure changes by the addition of the infalling observer, we can deduce bulk geometries with appropriate gravitational backreaction. This section and the next are summary of~\cite{Beni19, Beni19b}. This section focuses on the boundary perspective. 
\vspace{0.5\baselineskip}

\begin{figure}
\centering
 \includegraphics[width=0.45\textwidth]{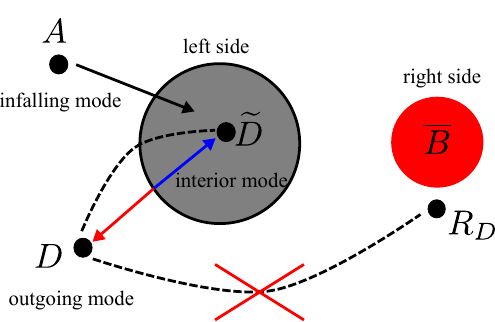} 
\caption{Backreaction from an infalling observer. The outgoing mode $D$ is entangled with a new mode $\widetilde{D}$ which is dynamically created by $A$ and has nothing to do with RHS.  
}
\label{fig-backreaction}
\end{figure}

For simplicity of discussion, we represent the black hole as a quantum system of $n$ qubits. As the initial state of the black hole, consider a generic maximally entangled state between the left side $B$ and the the right side $\overline{B}$:
\begin{align}
(I \otimes K)|\text{EPR}\rangle_{B\overline{B}} \qquad |\text{EPR}\rangle_{B\overline{B}}\equiv\frac{1}{\sqrt{d_{B}}}\sum_{j}|j\rangle_{B}\otimes |j\rangle_{\overline{B}} \label{eq:initial-K}
\end{align}
where $K$ is an arbitrary unitary operator. Imagine that some measurement probe (or an infalling observer) $A$ is dropped into the black hole at time $t=0$ as in Fig.~\ref{fig-backreaction}. Rather than keeping track of the outcomes for all the possible input states on $A$, it is convenient to append a reference system $\overline{A}$ which is entangled with the probe $A$ and forming EPR pairs $|\text{EPR}\rangle_{A\overline{A}}=\frac{1}{\sqrt{d_{A}}}\sum_{j}|j\rangle_{A}\otimes |j\rangle_{\overline{A}}$. 

After the time-evolution by some unitary dynamics $U$, the system evolves to
\begin{align}
|\Psi\rangle \equiv (U_{BA}\otimes I_{\overline{A}} \otimes K_{\overline{B}}) (|\text{EPR}\rangle_{B\overline{B}} |\text{EPR}\rangle_{A\overline{A}}) = \ \figbox{1.0}{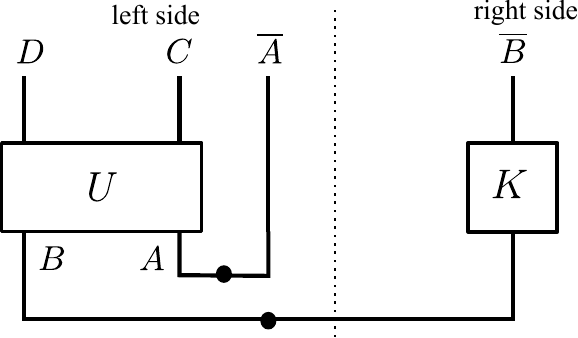} \label{eq:initial-state}
\end{align}
where $D$ is the outgoing mode and $C$ is the remaining black hole. Horizontal lines represent EPR pairs. Here black dots represent factor of $1/\sqrt{d_R}$ in a subsystem $R$ for proper normalization of $|\text{EPR}\rangle_{R\overline{R}}$. A dotted line divides the left and right sides of the black hole.
\vspace{0.5\baselineskip}

In the absence of the infalling observer $A$, the outgoing mode $D$ is entangled with some mode $R_{D}$ on the right-hand side $\overline{B}$. Here we are interested in finding what degrees of freedom $D$ is entangled with after adding the infalling observer $A$ to the system. 

Let us begin by briefly recalling the concept of quantum information scrambling. A quantum black hole delocalizes quantum information rapidly and its effect on boundary CFTs can be characterized by out-of-time order correlations (OTOCs). Let $V_{A}$ and $W_{D}$ be arbitrary traceless basis operators, such as Pauli or Majorana operators, supported on $A$ and $D$ respectively. Initially at $t=0$, we have $\langle V_{A}(0)W_{D}(0) V_{A}^{\dagger}(0)W_{D}^{\dagger}(0) \rangle=1$ when $V_{A}$ and $W_{D}$ do not overlap. Here we took the quantum state to be the maximally mixed state $\rho=\frac{1}{d}I$. After the scrambling time, on the other hand, OTOCs decay to small values:
\begin{align}
\langle  V_{A}(0)W_{D}(t) V_{A}^{\dagger}(0)W_{D}^{\dagger}(t)  \rangle \approx 0  \qquad t \gtrapprox t_{\text{scr}}
\end{align}
For more details on the definition of scrambling via OTOCs, see~\cite{Yoshida:2017aa}. 

The inclusion of the infalling observer has a drastic effect on the entanglement structure due to the scrambling dynamics of the black hole. The following statement was proven in~\cite{Beni18}. Suppose that the system is scrambling in the above sense and $d_A\gg d_D$~\footnote{Here $d_{R}$ represents the Hilbert space size of the subsystem $R$.
See~\cite{Beni19} for discussions on cases where the time separation is shorter than the scrambling time as well as a physical interpretation of the condition $d_A\gg d_D$. We recap these discussions in appendix~\ref{Sec:misc}.
}
. Then, the subsystems $D$ and $\overline{B}$ in $|\Psi\rangle$ are decoupled (not entangled):
\begin{align}
\rho_{D\overline{B}} \approx \rho_{D} \otimes \rho_{\overline{B}}  \label{eq:decoupling}.
\end{align}
where the error is $O\Big(\frac{d_{D}^2}{d_{A}^2}\Big)$. Hence, we arrive at the following conclusion:
\vspace{0.2\baselineskip}

\emph{
-- The outgoing mode $D$ is no longer entangled with the right-hand side $\overline{B}$. Instead $D$ is entangled with $C\overline{A}$. Thus the entangled partner mode is found exclusively on the left-hand side. 
}
\vspace{0.2\baselineskip}

An immediate, yet important implication of this conclusion is that the construction of the interior partner mode, which we shall denote by $\widetilde{D}$, is independent of the initial state of the black hole (or the unitary $K$ in Eq.~\eqref{eq:initial-K}). The same expression of the interior partner mode works for any initial state of the black hole, be it two-sided or one-sided since it is supported exclusively on the left-hand side without using degrees of freedom from $\overline{B}$. Hence, infalling observer's backreaction enables us to obtain \emph{state-independent} construction of interior partner operators~\footnote{It is worth emphasizing, however, that the construction depends on the initial state of the infalling observer, and thus is \emph{observer-dependent}. See~\cite{Beni19b} for details.}.

\vspace{0.5\baselineskip}

\begin{figure}
\centering
(a)\ \includegraphics[width=0.4\textwidth]{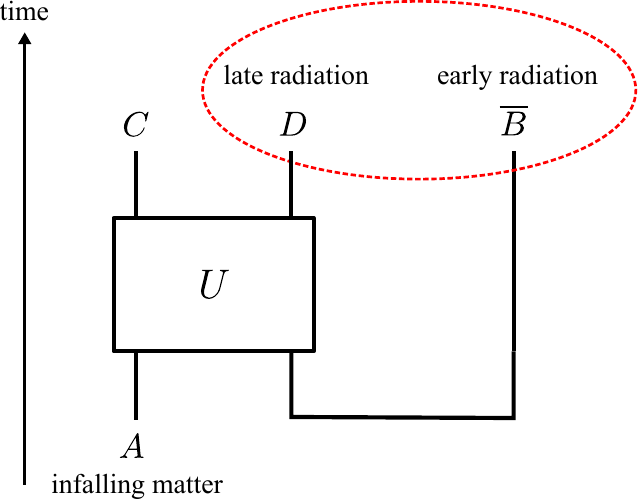} \ \ \ \
(b)\ \includegraphics[width=0.4\textwidth]{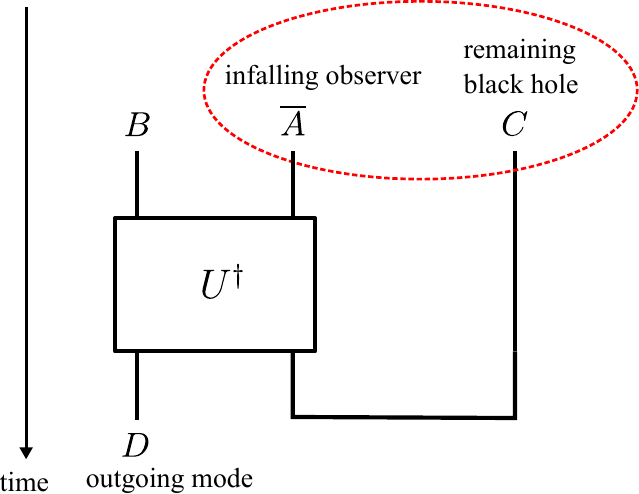} 
\caption{(a) The Hayden-Preskill recovery problem. (b) The interior reconstruction problem.  
}
\label{fig-relation}
\end{figure}

We briefly recap how to reconstruct the interior partner operators~\cite{Beni18, Beni19b}. Imagine that the dynamics of the black hole runs ``backward'' in time such that the outgoing mode $D$ returns back to the black hole at time $t$, and then the infalling observer $A$ (or its reference $\overline{A}$) is emitted back to the exterior at time $t=0$. It is convenient to represent the backward time-evolution with quantum circuit diagrams as in Fig.~\ref{fig-relation}. Our goal is to reconstruct the entangled partner mode of $D$ by having access to the remaining black hole $C$ at time $t$ and the emitted infalling observer $\overline{A}$ at time $t=0$. This interior reconstruction problem is mathematically equivalent to the Hayden-Preskill recovery problem which aims to reconstruct the infalling quantum state by having access to the early radiation and the late radiation. In the Hayden-Preskill recovery problem, $A$, $B$, and $D$ represent the infalling quantum state, the early radiation, and the late radiation respectively (Fig.~\ref{fig-relation}(a)). In the interior reconstruction problem, $D$, $C$, and $A$ play roles of the infalling quantum state, the early radiation, and the late radiation in the Hayden-Preskill problem respectively (Fig.~\ref{fig-relation}(b)). 

The upshot is that the problem of reconstructing interior partner operators can be interpreted as the Hayden-Preskill problem running backward in time. This observation enables one to obtain an explicit expression of interior operators by using recovery protocols for the Hayden-Preskill thought experiment. 

While being mathematically similar, the interior reconstruction problem differs from the Hayden-Preskill problem in important detail. In the Hayden-Preskill problem, the infalling quantum state can be reconstructed only for old black holes where $B$ is maximally entangled with the early radiation $\overline{B}$. Furthermore, since the recovery procedure accesses $\overline{B}$ and $D$, it requires non-locally coupling two sides of the black hole. On the other hand, in the interior reconstruction problem, one does not need access to $\overline{B}$ since the remaining black hole degrees of freedom $C$ plays the role of the early radiation. The reconstruction procedure accesses only to $C$ and $\overline{A}$ and thus does not require non-local coupling. As such, the interior reconstruction can be performed via one-sided quantum operations and works for young or one-sided black holes too. 

\vspace{0.5\baselineskip}

The important conclusion of this section is that the outgoing mode $D$ will be decoupled from the other side of the black hole (or the early radiation) $\overline{B}$ and a new partner mode $\widetilde{D}$ is dynamically created due to backreaction from the infalling observer. The newly created partner mode has nothing to do with the preexisting entanglement (or the wormhole) of the black hole. This clearly suggests the failure of the $\text{ER}=\text{EPR}$ approach in describing the experience of the infalling observer $A$. So far, our treatment mainly concerns the boundary perspective. The next task is to give a bulk interpretation of this phenomenon.

\section{Bulk Interpretation}\label{Sec:bulk}

We begin with the bulk interpretation of the emergence of the new partner mode $\widetilde{D}$. For simplicity of discussion, we will mainly focus on the two-sided AdS black hole which is in the thermofield double state while the core of the argument below will apply to generic maximally entangled black holes as well as one-sided black holes.

\begin{figure}
\centering
\includegraphics[width=0.4\textwidth]{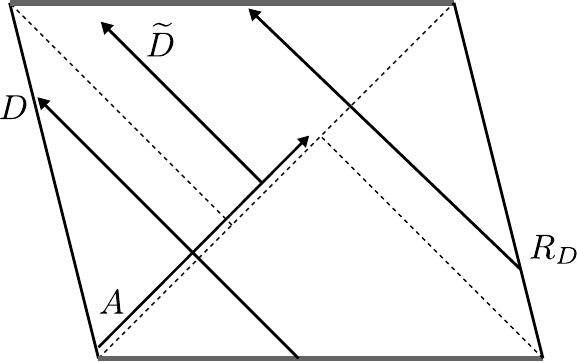} 
\caption{Bulk interpretation with a shockwave. 
}
\label{fig-bulk}
\end{figure}

In the absence of the infalling observer, the interior partner of the outgoing mode $D$ can be found on the other side of the black hole as depicted in Fig.~\ref{fig-bulk}. We shall denote this partner mode, constructed exclusively on the degrees of freedom on the right-hand side, by $R_{D}$. We now include the effect of the infalling observer $A$. The central idea is to treat the infalling observer $A$ as a gravitational shockwave and draw the backreacted geometry where the horizon is shifted as depicted in Fig.~\ref{fig-bulk}. If the time separation between the outgoing mode $D$ and the infalling observer $A$ is longer than or equal to the scrambling time, the near horizon geometry is significantly shifted along the trajectory of $A$. Namely there will be a new interior mode $\widetilde{D}$, \emph{distinct} from $R_{D}$, which can be found across the horizon as depicted in Fig.~\ref{fig-bulk}.

From the boundary analysis in the previous section, we already know that the new partner mode $\widetilde{D}$ can be constructed entirely on the left-hand side of the black hole. This fact can be also verified from the bulk perspective as follows. 
From the backreacted Penrose diagram in Fig.~\ref{fig-bulk}, we notice that any infalling signals from the right-hand side do not overlap with $\widetilde{D}$, and thus $\widetilde{D}$ is outside the causal influence of the right-hand side. This suggests that operators accounting for $\widetilde{D}$ must commute with any operators on the right-hand side, suggesting that $\widetilde{D}$ should be constructed entirely on the left-hand side~\footnote{Here we implicitly assumed that algebra of simple operators on RHS and their time-evolution generate the whole algebra of RHS by ignoring conservation laws.}. 

\vspace{0.5\baselineskip}

Next, let us discuss why the infalling observer $A$ will not see the signal $R_{D}$ from the other side. Recall that the new partner mode $\widetilde{D}$ is dynamically created as a result of interaction between $A$ and $D$ whose strength is quantified by the decay of OTOCs. This suggests that $\widetilde{D}$ is created once $A$ observes the outgoing mode $D
$ directly or indirectly outside the horizon. Namely, the rearrangement of the interior partner mode from $R_{D}$ to $\widetilde{D}$ has already occurred when $A$ crosses the horizon. The bulk picture, where the infalling observer $A$ herself is treated as a shockwave, suggests that $A$ will encounter the partner mode $\widetilde{D}$ shortly after crossing the horizon. This encounter will occur roughly at the same time and the location where trajectories of $A$ and $R_{D}$ appear to intersect in the unperturbed Penrose diagram in Fig.~\ref{fig-meeting}. The important difference is that $A$ will encounter $\widetilde{D}$, instead of $R_{D}$.

Some readers might think that $A$ would eventually encounter $R_{D}$ after her encounter with $\widetilde{D}$ as depicted in Fig.~\ref{fig-bulk}. Observe, however, that this encounter would happen almost at the singularity when the time separation between $A$ and $D$ is longer than or equal to the scrambling time. It is then not surprising if the naive bulk picture makes a false prediction near the singularity
~\footnote{One subtlety of this bulk interpretation is that we need to think that the infalling observer $A$ is located just ``above'' the shockwave so that she encounters $\widetilde{D}$ near the horizon and $R_{D}$ near the singularity. Then the bulk picture is consistent with predictions from the boundary quantum mechanics. One may interpret this mechanism in the following way. The infalling observer $A$ throws a few qubits a Planck time before she jumps into the black hole. This will guard herself against signals $R_{D}$. (Here we are not saying that $A$ actually needs to throw qubits beforehand. We claim that this mechanism occurs just by $A$'s jumping into a black hole).}. 

\vspace{0.5\baselineskip}

One successful application of the disentangling phenomenon by an infalling observer is the resolution of the firewall puzzle~\cite{Beni19}. Recall that  the outgoing mode $D$ was initially entangled with some degrees of freedom $R_{D}$ in $\overline{B}$. Assuming the smooth horizon, an infalling observer Alice would see an interior mode $\widetilde{D}$ which is entangled with the outgoing mode $D$. This, however, leads to a contradiction because $D$ is also entangled with $R_{D}$. If we would think that $D$ remains entangled with $R_{D}$, then $D$ and $\widetilde{D}$ would not be entangled, leading to possible high energy density at the horizon. 

The resolution of the firewall puzzle is immediate. When an infalling observer jumps into a black hole, the outgoing mode $D$ is disentangled from $R_{D}$ due to her own backreaction. She will be able to observe the interior mode $\widetilde{D}$ which is distinct from the original partner mode $R_{D}$~\footnote{
While we have already presented multiple counterarguments against the $\text{ER}=\text{EPR}$ approach, it is worth presenting another one to make doubly sure. In the $\text{ER}=\text{EPR}$ approach, it was argued that touching $R_{D}$ would create a deadly shockwave that would kill $A$ and prevents $A$ from seeing the interior partner mode. This is of course an incorrect statement because $A$ will not see $R_{D}$ at all. But let us suppose that this is the case to derive a contradiction. This would then suggest that, when $R_{D}$ was not perturbed, $A$ will be able to cross the horizon smoothly and sees the interior partner mode $R_{D}$. Recall, however, that the backreaction from $A$ disentangled $D$ from $R_{D}$. This contradicts the claim that $A$ crosses the horizon smoothly as it would suggest that $D$ and $R_{D}$ are entangled. Hence we conclude that the $\text{ER}=\text{EPR}$ proposal does not resolve the firewall puzzle.
}. We will revisit this resolution by strengthening the firewall puzzle from the perspective of the qECT thesis in Section~\ref{Sec:Discussion}.

\section{Universality}\label{Sec:Universality}

Based on the boundary quantum information theoretic analysis, we have argued that the inclusion of the infalling observer $A$ disentangles the outgoing mode $D$ from the other side of the black hole. From the bulk quantum gravity viewpoint, we have argued that this disentangling phenomena can be associated with the gravitational backreaction caused by the infalling observer $A$. 


From the infalling observer's perspective, however, the above conclusion might still sound counterintuitive. Imagine that you are the infalling observer who dives into the black hole. Upon encountering the outgoing radiation $D$ near the black hole horizon, you would have a freedom to choose to interact with $D$ or not. Suppose that you choose not to interact with $D$ at all. Without direct interaction between $A$ and $D$, the disentangling phenomena would not occur, and hence $D$ would remain entangled with the RHS. It then appears that our proposed mechanism for resolving the firewall puzzle would not work. Indeed, it would be rather strange if the presence (or absence) of the firewall depends on whether you measure $D$ or not.

The crucial insight in resolving this apparent puzzle is to observe that the disentangling phenomena, as well as the gravitational backreaction, occur universally, \emph{no matter how} the infalling observer $A$ jumps into a black hole. In the above argument, we implicitly assumed that, by choosing not to measure $D$ directly, $A$ and $D$ would remain non-interacting. There are, however, indirect interactions between $A$ and $D$, mediated by the effect of quantum gravity, which cannot be captured locally by effective quantum field theory in the infalling observer's local frame. In this section, we elaborate on this observation from both the boundary and bulk perspectives.


\vspace{0.5\baselineskip}

Let us begin with the boundary quantum mechanical viewpoint. Regardless of how the infalling observer is introduced, what is important for us is whether OTOCs between $A$ and $D$ will eventually decay or not. One might wonder whether an observer can be introduced in a way which introduces no backreaction or decay of OTOCs. We speculate that this is impossible. The disentangling phenomena relies on decay of OTOCs which is a rather generic feature of interacting quantum many-body systems. It is in principle possible that the value of OTOCs returns to $O(1)$ after a very long time (of the order the quantum recurrence time) in a strongly coupled quantum system. But in time scales which are relevant to the experience of the infalling observer, OTOCs decay monotonically to a small stationary value and then oscillates around it. Such a conclusion can be mathematically proven by utilizing the Eigenstate Thermalization Hypothesis~\cite{Huang:2019aa}. Hence the insertion of $A$ will eventually lead to the disentangling phenomena.


From the bulk quantum gravitational perspective, any perturbation inserted at the AdS boundary or the asymptotic infinity will always lead to a significant gravitational backreaction near the black hole horizon, no matter how small the initial perturbations are. What amplifies the effect of the backreaction is the blueshift caused by the black hole horizon which universally accelerates the infalling quanta at the rate determined by its surface gravity. As long as the infalling object carries a small but finite positive energy (including its rest mass), it eventually introduces significant gravitational backreaction when approaching the black hole horizon. Hence, the disentangling phenomena always occurs no matter how the infalling observer jumps into the black hole.

One might wonder if adiabatically inserting an infalling observer and moving her slowly to the black hole horizon would avoid the gravitational backreaction. While we are not able to compute its effect explicitly, we speculate that such adiabatic insertion will necessarily introduce significant gravitational backreaction~\footnote{It will be interesting to study the modular Berry connection~\cite{Czech:2018aa} in the present context.}. For one thing, in evaluating OTOCs $\langle O_{A}(0)O_{D}(t)O_{A}(0)O_{D}(t) \rangle$, one needs to design $O_{A}$ carefully so that it creates a slowly moving excitation which would somehow stop and hover near the black hole horizon. It is typically the case that such complex operators lead to further decay of OTOCs as $O_{A}$ would have higher weights of support. Also, slowing down $A$ would require us to counteract the acceleration from the black hole by some means which will be enough to induce the gravitational backreaction and the disentangling phenomena.

\vspace{0.5\baselineskip}

Several comments are in order. The interaction between $A$ and $D$ is indirect and cannot be seen in the infalling observer's local frame. Its effect can be verified only in a global perspective at the AdS boundary or the asymptotic infinity in the form of decay of OTOCs. 



Based on the aforementioned observations, it is worth revisiting the underlying assumptions that had led to the AMPS puzzle; 1) a smooth horizon (no drama), 2) quantum mechanics (unitarity, locality and monogamy of entanglement), and 3) validity of bulk effective field theory. Authors of~\cite{Almheiri13} argued that these three assumptions are not compatible with each other. In our proposed scenario, all the three assumptions remain valid. Yet, the puzzle is avoided due to the subtlety concerning the interpretation of the prediction from the bulk effective field theory. While, in the infalling observer's frame, the bulk effective field theory is perfectly valid, the partner mode $D$ has been already rearranged to a new mode $D$ when $A$ is included to the system by choosing the observer's frame. It was wrong to assume that the interior partner mode is unique, see~\cite{Beni19b} for further details.

The disentangling phenomena occurs due to the positivity of the energy associated with the infalling quanta. While it is prohibited to create a negative energy perturbation from simple one-sided action, by coupling two sides of the black holes in a suitable manner, one is able to send negative energy to the black hole in principle. Examples of such perturbations include the traversable wormhole phenomena and the Hayden-Preskill recovery protocol. These protocols have an effect of saving the infalling observer $A$ from falling into the black hole, and hence, preventing her from seeing the interior partner mode. In a sense, such negative energy perturbations behave like a firewall by preventing someone from crossing the black hole horizon. Its effect, however, is more peaceful as it extracts the infalling observer from the fall into the black hole, as opposed to the conventional picture of a firewall which would kill the infalling observer. See~\cite{Beni19} for further discussions along this line.

\vspace{0.5\baselineskip}

In summary, we claim that any attempt to see the black hole interior necessarily introduces significant gravitational backreaction. This leads to disentangling the outgoing mode from the other side of the black hole. This phenomena occurs universally and originates from the existence of the black hole horizon. Furthermore, this effect creates a new partner mode $D$ in a way which depends on how the infalling observer is introduced to the system. While the bulk effective field theory remains valid in the infalling observer's local frame, its interpretation is subtle as the construction of the partner mode in the language of the boundary quantum mechanics dynamically changes.



\section{Measurement}\label{Sec:Measurement}

We have seen that any unitary operation from the other side of the black hole will not affect the experience of the infalling observer. In this section, we discuss whether a measurement may influence the infalling observer or not. 

There are two different formulations of measurements. The first way, referred to as the Copenhagen interpretation, is to accept the collapse of wavefunctions after performing a measurement. In this interpretation, an observer is introduced to the quantum system of interest and performs a measurement which applies projections according to the Born rule. This replaces the system with a density matrix, and hence, in this viewpoint, the measurement is an irreversible process. The second way, often referred to as the Everett relative state interpretation, is to view a measurement as a unitary process that entangles the system with apparatus and the environment. In this viewpoint, the measurement is a process which can be reversed in principle.

In the Everett interpretation, the whole system evolves unitarily, so our analysis from previous sections suggest that a measurement on the other side will not affect the infalling observer. In the Copenhagen interpretation, however, the problem appears to be more subtle as a measurement destroys entanglement between two sides of the black hole. In a series of papers~\cite{Susskind14, Susskind:2016aa}, Susskind argued that a measurement will influence the infalling observer since the wormhole connecting two sides will be terminated via a measurement in the Copenhagen interpretation by invoking the $\text{ER}=\text{EPR}$ proposal. 

Here we arrive at a different conclusion. Recall that the construction of the interior partner operator is independent of the initial state of the black hole and is supported exclusively on the left hand side (the infalling observer's side) of the black hole, as discussed in section~\ref{Sec:Backreaction}. This implies that any quantum operation, including measurements, on the other side of the black hole will not affect the infalling observer. Even if one projects the other side to some wavefunction, the partner mode $\widetilde{D}$ remains entangled with the outgoing mode $D$, suggesting that a measurement will not affect the infalling observer in the Copenhagen interpretation. The bulk picture is similar to the one from section~\ref{Sec:bulk}.

Hence, we conclude that, in both the Copenhagen and Everett interpretations, a measurement on the other side of the black hole will not influence the infalling observer. 

\section{Resolution}\label{Sec:Resolution}

Let us finally return to the black hole complexity puzzle. We begin by arguing that Susskind's protocol for measuring $\frac{d\mathcal{C}}{dt}$ cannot be performed. In his protocol, an infalling observer from one side jumps into the black hole and see if she will be hit by a shockwave from the other side or not. As is already mentioned, it is strange to expect that the infalling observer could see the shockwave when two sides of the black hole are not coupled. The key to resolve this misunderstanding was to consider the effect of backreaction by an infalling observer as in Fig.~\ref{fig-bulk}. Assume that the shockwave was created by exciting the boundary mode $R_{D}$. Since this mode $R_{D}$ was initially entangled with $D$, the infalling observer would expect to be hit by the entangled partner mode of $D$ behind the horizon. Due to her own backreaction, however, $D$ will no longer be entangled with $R_{D}$ once the infalling observer jumps into the black hole. Inside the black hole, she will just see the interior mode $\widetilde{D}$ which is distinct from $R_{D}$. Hence the infalling observer will not be able to see the shockwave from the other side. 

Here we assumed that two sides of the black hole are not coupled. For two signals/observers from opposite sides to meet inside the black hole, two boundaries need to be coupled appropriately. By using the traversable wormhole phenomenon~\cite{Gao:2017aa} or the Hayden-Preskill decoding protocol~\cite{Yoshida:2017aa}, two observers may be able to see each other and then travel to the other sides of the black hole. To implement these protocols, however, one needs to reduce the boundary CFT wavefunctions back to low complexity states (such as the TFD state), so there is no shortcut in these approaches. It is unclear to us if there would be a simpler way to make two observers meet inside the black hole. To influence two decoupled degrees of freedom deep in the bulk, it is natural to expect that high complexity quantum operations are needed on boundary CFTs. 
\vspace{0.5\baselineskip}

\begin{figure}
\centering
(a)\ \includegraphics[width=0.23\textwidth]{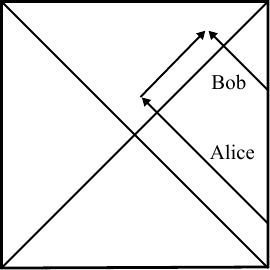} \ \ \ \
(b)\includegraphics[width=0.23\textwidth]{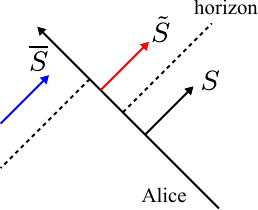} 
\caption{(a) A setup by Hayden and Preskill. Alice sends a signal to Bob. (b) A refinement of the Hayden-Preskill argument. Alice's signal does not reach Bob due to backreaction. 
}
\label{fig-same}
\end{figure}

Next, let us turn to BFV's protocol. Here we shall focus on a certain subroutine of their protocol. BFV's protocol requires a pair of observers who send signals to each other inside the black hole. Since two observers from opposite sides cannot communicate as discussed above, they should come from the same side of the black hole. Let us imagine that one observer (Alice) jumps into the black hole, and then the other observer (Bob) jumps later. Upon crossing the horizon, Alice sends a signal to Bob. Can they communicate? Hayden and Preskill asked exactly the same question in~\cite{Hayden07}. Their argument goes as follows. If the time separation $\Delta t$ between Alice and Bob is longer than the scrambling time, Alice needs a large amount of energy, possibly larger than the Planck energy, to send a signal to Bob. Otherwise, Bob will reach singularity before Alice's message arrives. Hence two observers will not be able to communicate with each other unless they introduce drastic backreaction to the underlying geometry. This suggests that there seems to be a fundamental upper bound on efficiently measurable volume or length near the horizon. 

Here we would like to present a modern perspective on Hayden-Preskill's observation by considering the effect of backreaction by Alice. Suppose that Alice was planning to send a signal by using the mode $\overline{S}$ behind the horizon, which is the partner mode of $S$ (Fig.~\ref{fig-same}(b)). Once Alice crosses the horizon, a new partner mode $\tilde{S}$ is dynamically created. Hence Alice cannot send a signal to Bob by using the original partner mode $\overline{S}$. To avoid backreaction, Alice would need to use the interior mode $\overline{S}$ which is not close to the horizon. However, this means that the time separation $\Delta t$ between Alice and Bob is significantly shorter than the scrambling time. 

While we have arrived at the same conclusion as Hayden and Preskill's, there is a subtle (but important!) refinement in our argument. The reason why Alice cannot communicate with Bob is that Alice has intersected with the mode $S$, and as a result, $\overline{S}$ is disentangled from $S$. Namely, the disentangling phenomenon occurs automatically, regardless of whether Alice attempts to send a signal to Bob with trans-Planckian energy or not.
\vspace{0.5\baselineskip}

Finally, we critically comment on Susskind's proposal for a resolution of the complexity puzzle~\cite{Susskind20}. Susskind proposed that the qECT thesis may be violated behind the black hole horizon, but the violation should not be communicated to the outside efficiently. This proposal is in accord with the black hole complementarity; one chooses not to be bothered by the violation of the qECT thesis if the violation cannot be ever found out. As we have already demonstrated in the main body of this note, backreaction from infalling observers invalidate any shortcuts to measuring the volume of the wormhole. Hence we do not need to invoke the black hole complementarity to extract the qECT thesis and resolve the puzzle. The effect of gravitational backreaction prevents the violation from happening. It is wrong to say that the violation is invisible; the violation simply did not happen~\footnote{Hayden and Preskill considered a thought experiment which consists of two components. The first component concerns the information recovery from an old black hole by an outside observer. This appears to create two copies of the identical quantum state inside and outside the black hole. The black hole complementarity asserts that it is fine to have a violation of the no-cloning theorem as long as it cannot be verified by any observer. Recent progress on scrambling dynamics of black holes, however, suggest that the black hole complementarity is not a correct resolution of the cloning puzzle. The modern interpretation is that the Hayden-Preskill recovery procedure introduces significant backreaction which pulls the quantum state in the interior to the exterior by shifting the location of the horizon~\cite{Gao:2017aa}. It is wrong to say that there are two copies of the same quantum state which cannot be verified by a single observer. There is always a single copy of the state only. The second component concerns a setup where Alice and Bob try to communicate with each other inside the black hole in an attempt to verify a violation of the no-cloning theorem. We have already discussed this case in the main body of the note and demonstrated that the black hole complementarity is not a correct approach.}.

Another counterargument can be obtained by actually saving the infalling observer from a black hole, as we have already discussed in section~\ref{Sec:ER=EPR}. Recall that an infalling observer in a two-sided black hole can be extracted from the black hole by quantum operations which act solely on one side of the black hole in principle~\cite{Beni19}. The quantum circuit complexity of quantum operations is independent of the complexity of the initial state (\emph{i.e.} the wormhole volume $\mathcal{V}$ of the two-sided black hole). Hence the infalling observer can report her experience after performing quantum operations with complexity much smaller than $\mathcal{V}$. This does not invalidate the qECT thesis; the rescued observer will not give us any useful information on the wormhole volume $\mathcal{V}$ or $\frac{d\mathcal{V}}{dt}$ as she was not hit by the shockwave. 

\section{Extended firewall puzzle}\label{Sec:extend}

Motivated by the black hole complexity puzzle, we would like to point out another challenge to the qECT thesis. The challenge concerns a certain puzzle on the decoding complexity (quantum circuit complexity for reconstruction) of interior partner modes. We view this as an extension of the firewall puzzle from the computational complexity perspective. In this section, we formulate the extended firewall puzzle and present speculation about its potential resolution.

\vspace{0.5\baselineskip}

As briefly reviewed in section~\ref{Sec:bulk}, the gravitational backreaction by an infalling observer provides a potential resolution of the firewall puzzle since the outgoing mode $D$ is decoupled from the early radiation $\overline{B}$ due to scrambling dynamics of a black hole and the new partner mode $\widetilde{D}$ is dynamically created. As such, the fact that the infalling observer can cross a smooth horizon and see interior partner modes does not lead to any inconsistency from the perspective of the monogamy of quantum entanglement~\cite{Beni19}. 

From the computational complexity perspective, however, we find some trouble in this explanation. From the viewpoint of the boundary observer, constructing the expression of interior partner operators appears to be a rather complex task. This is indeed the case especially when the Hilbert space size accounting for interior modes is large and OTOCs have decayed to asymptotic values at late times (when the time separation $\Delta t$ between the infalling observer $A$ and the outgoing mode $D$ becomes large). The best-known algorithm for interior reconstruction at such late times uses the Grover search algorithm whose runtime is proportional to the Hilbert space size of the interior modes (hence is exponential in the number of qubits)~\cite{Yoshida:2017aa}. As such, the decoding complexity of interior partner modes is high for the boundary observer. Yet, the situation appears to be rather different for the bulk observer. Namely, the infalling observer can see the interior Hawking modes by simply jumping into a black hole and crossing the smooth horizon without doing any difficult quantum computation. 

Hence we seem to find that the decoding complexity of interior partner modes is high (low) for boundary (bulk) observers respectively. This leads to an apparent violation of the qECT thesis. We would like to refer to this as the extended firewall puzzle. 

\vspace{0.5\baselineskip}

Below we present a possible scenario for resolving the extended firewall puzzle. When the time separation $\Delta t$ between the infalling observer and the outgoing mode is longer than the scrambling time $t_{\text{scr}}$, the infalling observer will encounter the interior partner degrees of freedom near the horizon, possibly at the Planck length distance (or less) away from the horizon. Such interior modes at short distance scale should not be visible to an infalling observer since these would be realized as a part of the geometry itself, instead of matter fields propagating freely on the geometry, due to some short distance quantum gravity effect. Hence it is reasonable to assume that interior partners at short distance scales are computationally difficult to decode for the infalling observer too. On the other hand, when the time separation $\Delta t$ is shorter than $t_{\text{scr}}$, one may utilize decoding protocols based on traversable wormhole effects to efficiently construct interior partner operators as in~\cite{Gao:2017aa}. These protocols utilize the fact that black holes are the fastest scrambler with the Lyapunov exponent $\frac{2\pi}{\beta}$ and work reliably only when a black hole scramble quantum information coherently until $\Delta t \lessapprox t_{\text{scr}}$. In the bulk, this corresponds to the fact that the infalling observer will encounter these interior partner modes away from the horizon, and can observe them easily as conventional matter fields. Hence, the qECT thesis will remain valid.

\vspace{0.5\baselineskip}

A resolution of the original firewall puzzle, in a sense of the monogamy of quantum entanglement, relied on the fact that a black hole scrambles quantum information. This resolution, however, is unsatisfactory as it did not utilize the fact that a black hole scrambles quantum information in the fastest possible manner. After all, a piece of burning wood will scramble quantum information at late times too! The extended firewall puzzle, strengthened by the qECT thesis and the computational complexity, naturally calls for the very fact that a black hole is the fastest scrambler. Namely, the separation of the decoding complexity of interior operators before and after the scrambling time results from the fast scrambling nature of the black hole. On the bulk, the separation of decoding complexity to different physical objects; the matter and the geometry. We hope to dwell on this scenario more in a separate work.

\section{Summary and discussions}\label{Sec:Discussion}

\subsection{Main results}

We summarize the main results of this note. 

\begin{itemize}
\item We have presented a possible resolution of the black hole complexity puzzle by considering the effect of gravitational backreaction by bulk observers. Our observation suggests that there is no shortcut in measuring the volume of the wormhole, and thus the qECT thesis remains valid. 

\item We have revisited the thought experiment by Hayden and Preskill which asked whether two infalling observers from the same side of a black hole can communicate with each other or not. We have presented a refinement of their argument (but with the same conclusion) by examining gravitational backreaction without invoking the black hole complementarity. 

\item We have argued that two observers from opposite sides of the black hole will not encounter each other inside the black hole. We have then discussed why the $\text{ER}=\text{EPR}$ approach does not revolve the firewall puzzle. Finally, we have reviewed an alternative resolution of the firewall puzzle by considering the effect of perturbation from an infalling observer. 

\item We have formulated the extended firewall puzzle by pointing out that the decoding complexity of interior operators appears to be high (low) for boundary (bulk) observers respectively. We have then presented speculation about its potential resolution by utilizing the fact that a black hole is the fastest scrambler. 
\end{itemize}

\subsection{Discussions}

We conclude this note with various discussions.

\subsubsection*{-- Entanglement entropy} 

A certain puzzle similar to the BFV's has been recently pointed out, concerning the measurability of entanglement entropy~\cite{Gheorghiu20}. Perhaps appropriate considerations of gravitational backreaction from bulk observers may provide a resolution of this puzzle~\footnote{\cite{Bao:2015aa} presents a certain argument concerning the measurability of entanglement entropy. We are however unsure about its relevance to~\cite{Gheorghiu20} or our work.}.

\subsubsection*{-- Diverging Hilbert space} 

Our results also provide a potential resolution of the unboundedness puzzle concerning the diverging Hilbert space dimension of the black hole interior (in a sense that the interior volume is diverging even after short-distance regularization). Our argument suggests that the apparent volume-law degrees of freedom can be regulated to give an area-law entropy dynamically due to backreaction. We speculate that similar reasoning may resolve the puzzle concerning Wheeler's ``bag of gold'' spacetime which also concerns a possible unboundedness of the black hole Hilbert space~\cite{Marolf:2009aa}. It is an interesting future problem to make this idea more quantitative. 

\subsubsection*{-- de Sitter space}

Conventional dS/CFT correspondence ends up with non-unitary CFTs at the spatial infinity~\cite{Strominger:2001aa}. Since non-unitary time-evolution have more computational power than unitary time-evolutions in general, such approaches do not seem to be fruitful from the perspective of the qECT thesis. It will be interesting to ask if the qECT remains valid for quantum gravity in de Sitter space. 

\subsubsection*{-- Halting problem}

A certain class of spacetimes, including the Kerr and the Reissner-Nordstr{\"o}m metrics, has an intriguing, yet puzzling, feature that an observer could keep traveling to new universes forever. Such a spacetime could be used to decide the halting problem, which is known to be undecidable by a Turing machine (be it classical or quantum)~\cite{Etesi:2002aa}. This is of course in an obvious tension with the qECT thesis. We speculate that proper consideration of perturbation and gravitational backreaction from an observer will resolve this puzzle.

\subsubsection*{-- Effective field theory}
 
Recently Herman Verlinde proposed an interesting argument on another possible loophole in the $\text{ER}=\text{EPR}$ proposal~\cite{H_Verlinde20}. The idea is that bulk effective theories emerge after projecting the system into a certain codeword subspace, and the preexisting quantum entanglement between two sides, as well as the quantum circuit complexity, do not play essential roles for the wormhole geometry. We believe that our treatment of backreaction by infalling observers provides explicit construction of such a projection operator. 

\subsubsection*{-- Holographic dictionary}
 
BFV demonstrated the complexity puzzle in a slightly different manner by considering the complexity of the bulk-boundary dictionary and arrived at the following conclusion; 1) the qECT thesis is violated or 2) the complexity of the holographic dictionary is high. In our argument, this no-go result has been avoided by the fact that the dictionary is observer-dependent and thus is not invariant. Indeed, the expression of the interior partner mode dynamically changes due to backreaction and depends on the initial state of the infalling observer.

\subsubsection*{-- Multiple copies}

Our argument does not address the computational power of using multiple copies. For instance, the Grover search protocol from~\cite{Yoshida:2017aa} is closely related to higher R\'{e}nyi entropies and multiple shockwave geometries which can be interpreted as properties of multiple copies. We speculate that the BFV protocol may be improved by incorporating a similar protocol.

\subsubsection*{-- Quantum error-correction}
 
Can we introduce bulk observers in a way that does not introduce backreaction? By applying certain quantum operations on the boundary, it is possible to cancel backreaction from bulk observers and save the naive bulk description from breaking down. Indeed, the traversable wormhole phenomenon and the Hayden-Preskill decoding protocol are examples of such operations. In the language of quantum error-correction, a naive bulk description, which mistakingly ignore backreaction, is valid only inside some codeword subspace. Backreaction by bulk observers can be viewed as ``errors'' to a quantum error-correcting code which brings the system outside the codeword subspace. By performing suitable quantum error-correction which cancels backreaction, one can keep the system inside the codeword subspace where the bulk effective description remains valid.

\vspace{0.5\baselineskip}

\subsection*{Acknowledgment}

I thank Adam Bouland and Nick Hunter-Jones for discussions (v1). I thank Hrant Gharibyan, Junyu Liu, Geoff Penington, and Douglas Stanford for discussions (v2). I thank the anonymous referee for suggesting to be explicit about the negation of the $\text{ER}=\text{EPR}$ proposal (v3). Research at the Perimeter Institute is supported by the Government of Canada through Innovation, Science and Economic Development Canada and by the Province of Ontario through the Ministry of Economic Development, Job Creation and Trade.

\appendix

\section{Miscellaneous matters}\label{Sec:misc}

In this appendix, we provide justifications of two assumptions in the decoupling theorem. This appendix is a reprint from~\cite{Beni19b}. 

\vspace{0.5\baselineskip}

We argued that two signals from opposite sides of a black hole will not encounter each other inside the black hole. The core of our argument was that a new interior mode $\widetilde{D}$ is dynamically created as a result of scrambling dynamics of a black hole and the bulk effective descriptions break down due to gravitational backreaction from an infalling observer. This decoupling mechanism relies on two technical assumptions; 1) the time separation $\Delta t$ between an observer $A$ and the outgoing mode $D$ satisfies $\Delta t \geq t_{\text{scr}}$ and  2) $d_{A}\gtrapprox d_{D}$ where $d_{A}, d_{D}$ are Hilbert space sizes of $A, D$. 

\vspace{0.5\baselineskip}

Below we discuss the cases which do not satisfy these assumptions. We begin with the cases with $\Delta t \leq t_{\text{scr}}$. 

\begin{enumerate}[a)]
\item \emph{Singularity:} If the time separation $\Delta t$ is much shorter than the scrambling time $\Delta t \ll t _{\text{scr}}$, the trajectories of $A$ and $R_{D}$ will intersect near the black hole singularity. It is then not surprising if bulk effective descriptions become invalid in the proximity of the singularity. This observation, however, does not provide a satisfactory justification for intermediate time scales $t_{\text{th}}\ll \Delta t \lessapprox t_{\text{scr}}$ where $t_{\text{th}}$ is the thermalization time. 

\item \emph{Entanglement quality:} If $\Delta t \lessapprox t_{\text{scr}}$, an infalling observer will interact with the outgoing mode $D$ away from the horizon.
The proper temperature near the Rindler horizon is given by $T = \frac{1}{2\pi \rho}$ where $\rho$ is the proper distance from the horizon. This suggests that, as one moves away from the horizon, the density of thermal entropy becomes small and the quality of quantum entanglement, which an infalling observer may be able to detect via free fall, may deteriorate. Hence, there is no serious inconsistency with our conclusions even if the decoupling phenomenon is weak.

\item \emph{Measurement strength:} Recall that the decoupling phenomenon occurs as a result of the interaction between the observer $A$ and the outgoing mode $D$. Its strength is quantified by the decay of OTOCs. This suggests that the decoupling, as well as the creation of a new partner mode $\widetilde{D}$, occurs as much as the observer $A$ measures $D$. Then, even if $D$ and $\widetilde{D}$ are not perfectly entangled, there is no contradiction with the monogamy of entanglement since $A$ is observing $D$ (and its partner $\widetilde{D}$) only weakly. 

If $A$ wishes to measure $D$ more directly or strongly, she may increase the size of herself ($d_{A}$) to enhance gravitational scattering or perform actual physical measurement on $D$ upon encountering it. We expect that such strong interactions or direct measurements will lead to significant decay of OTOCs and create $\widetilde{D}$ which is nearly perfectly entangled with $D$.
\end{enumerate}

\vspace{0.5\baselineskip}

Next, let us justify the condition of $d_{A}\gtrapprox d_{D}$ by presenting three observations.

\begin{enumerate}[a)]
\item[d)] \emph{AMPS:} For an application to the firewall puzzle, namely the AMPS thought experiment, we can justify the requirement. In the thought experiment, the outside observer distills $R_{D}$ and hand it to the infalling observer $A$ who jumps into the black hole to verify the entanglement between $D$ and $\widetilde{D}$. This effectively realizes a situation with $d_{A}= d_{D}$. See~\cite{Beni19b} for details.
\item[e)] \emph{Metaphysical explanation:} To experience some physics, the infalling observer $A$ herself should carry some amount of entropies, at least as much as the objects $D$ she is going to measure. While being metaphysical, this explanation can be further justified from the aforementioned observation c) which suggests the decoupling occurs as much as the observer $A$ measures $D$. 

\item[f)] \emph{QFT:}
One may interpret $A$ and $D$ as infalling and outgoing modes of low energy QFT on the LHS wedge respectively while $B$ and $C$ can be viewed as all the other high-energy degrees of freedom associated with the future and past horizons respectively. It is thus natural to assume $d_{A} = d_{D}$ in a non-evaporating black hole as in the AdS space. (For an evaporating black hole, one can model the dynamics by taking $d_{D}$ to be slightly larger than $d_{A}$. The $\frac{d_{D}}{d_{A}}$ portion of interior operators becomes state-dependent and leaks out from the black hole, leading to the Page curve behavior). Here $A$ and $D$ are DOFs which an outside observer can easily see.

On the other hand, the infalling observer, who travels nearly at the speed of light, will propagate along with the infalling mode $A$. Hence it is natural to expect that the initial state of $A$ herself becomes irrelevant as long as the initial state is chosen from typical states in the low energy subspace of the QFT. Regardless of the initial state of $A$, she should be able to see the same bulk effective QFT description near the horizon.  This is indeed the case as the creation of $\tilde{D}$ occurs for any initial state of $A$. However, it is worth emphasizing that the map from the bulk QFT to the boundary is dependent on the initial state of $A$. Here the infalling observer will have access to the outdoing mode $D$ and the interior mode $\widetilde{D}$ instead of $A$. 

\end{enumerate}

\providecommand{\href}[2]{#2}\begingroup\raggedright\endgroup

%
\end{document}